\documentclass[letterpaper,titlepage,11pt]{article}
\usepackage{latexsym,amssymb,amstext,amsmath}
\usepackage{slashed}
\usepackage{mathrsfs}
\usepackage{color}
\usepackage{enumerate}
\usepackage{graphicx}
\usepackage{tikz}
\usepackage{etex}
\usepackage{latexsym}
\usepackage{cite}
\allowdisplaybreaks

\usepackage[
colorlinks=true,
linkcolor=blue,
urlcolor=red,
filecolor=green,
citecolor=red,
pdfstartview=FitV,
pdftitle={},
pdfauthor={Mehmet Ozkan},
pdfsubject={},
pdfkeywords={},
pdfpagemode=None,
bookmarksopen=true
]{hyperref}

\newcommand{\bea}{\setlength\arraycolsep{2pt} \begin{eqnarray}}
\newcommand{\eea}{\end{eqnarray}}
\newcommand{\nn}{\nonumber}

\usepackage{hyperref}

\newsavebox{\uuunit}
\sbox{\uuunit}
{\setlength{\unitlength}{0.825em}
	\begin{picture}(0.6,0.7)
	\thinlines
	\put(0,0){\line(1,0){0.5}}
	\put(0.15,0){\line(0,1){0.7}}
	\put(0.35,0){\line(0,1){0.8}}
	\multiput(0.3,0.8)(-0.04,-0.02){12}{\rule{0.5pt}{0.5pt}}
	\end {picture}}

\def\be{\begin{equation}}
\def\ee{\end{equation}}
\def\ba{\begin{array}}
\def\ea{\end{array}}
\def\bea{\begin{eqnarray}}
\def\eea{\end{eqnarray}}
\def\bd{\begin{displaymath}}
\def\ed{\end{displaymath}}

\def\nn{\nonumber}

\def\a{\alpha}
\def\b{\beta}
\def\g{\gamma}

\def\d{\delta}

\def\e{\epsilon}
\def\ve{\varepsilon}

\def\p{\psi}
\def\P{\Psi}

\def\l{\lambda}

\def\m{\mu}
\def\n{\nu}
\def\r{\rho}

\def\t{\tau}

\def\o{\omega}
\def\O{\Omega}

\def\nn{\nonumber}

\def\cN{\mathcal{N}}

\makeatletter
\@addtoreset{equation}{section}
\makeatother

\begin{document}
%
\begin{titlepage}

\bigskip

\begin{center}
{\LARGE\bfseries Three-Dimensional Extended Lifshitz, Schr\"odinger and Newton-Hooke Supergravity}
\\[10mm]

\textbf{Nese Ozdemir, Mehmet Ozkan and  Utku Zorba}\\[5mm]
\vskip 25pt
%
%
{\em   \hskip -.1truecm Department of Physics, Istanbul Technical University,  \\
	Maslak 34469 Istanbul, Turkey  \vskip 5pt }
{email: {\tt nozdemir@itu.edu.tr, ozkanmehm@itu.edu.tr, zorba@itu.edu.tr}}

\end{center}

\vspace{3ex}

\begin{center}
{\bfseries Abstract}
\end{center}
\begin{quotation}

We provide a systematic analysis of three-dimensional $\cN=2$ extended Bargmann superalgebra and its Newton-Hooke, Lifshitz and Schr\"odinger extensions. These algebras admit invariant non-degenerate bi-linear forms which we utilized to construct corresponding Chern-Simons supergravity actions.

\end{quotation}

\vfill

\flushleft{\today}
\end{titlepage}
\setcounter{page}{1}
\tableofcontents

\newpage

\section{Introduction}{\label{Intro}}

Newton-Cartan gravity is the geometric reformulation of the Newtonian gravity that is described by a degenerate spatial metric $h^{\m\n}$, a temporal vielbein $\t_\m$ and a $U(1)$ connection $m_\m$\footnote{In this paper, the Greek indices $(\m,\n,\ldots)$ refer to the coordinate frame and labels all spacetime coordinates with $x = (t, \vec{x})$ while the Latin alphabet letters $(a,b,\ldots)$ refer to the spatial local Galilean frame.}. This geometric setup of Newtonian gravity is realized based on two principles: (i) The geodesic equation is equivalent to the classical equation of motion of a massive particle, and (ii) The only non-vanishing component of the Riemann tensor gives rise to the Poisson equation of Newtonian gravity \cite{Cartan1,Cartan2}. If the Riemann tensor satisfies the Trautman \cite{Trautman} and the Ehlers \cite{Ehlers} conditions, then we are led to the so-called Newtonian connection that is the only non-vanishing component of the connection is given by $\Gamma^a_{00} = \delta^{ab} \partial_b \phi$, satisfying (i). Consequently, the only non-vanishing component of the Riemann tensor becomes the desired Poisson's equation for Newtonian gravity
\bea
R^a{}_{0a0} (\Gamma)= \nabla^2 \phi = 4 \pi G \rho \,,
\label{NewtonCartan}
\eea
where $G$ is Newton's constant and $\rho$ is the mass density. An alternative derivation of this geometric formulation can be achieved by gauging the underlying symmetry algebra that is the Bargmann algebra \cite{EricPanda}. The Bargmann algebra consists of the generators of spatial rotations $J_{ab}$, spatial translations $P_a$, Galilean boosts $G_a$, time translations $H$ and a central charge $M$, corresponding to particle mass. In the gauging procedure, one first associates each generator with a gauge field 
\bea
J_{ab} \rightarrow \o_\m{}^{ab}, \quad P_a \rightarrow e_\m{}^a, \quad G_a \rightarrow \o_\m{}^a, \quad H \rightarrow \t_\m, \quad M \rightarrow m_\m \,,
\eea
and imposes constraints on the curvatures of $\t_\m, e_\m{}^a$ and $m_\m$
which connect the gauge transformations to general coordinate transformations and solve $\o_\m{}^{ab}$ and $\o_\m{}^a$ in terms of $e_\m{}^a, \t_\m$ and $m_\m$, leaving the triplet $(e_\m{}^a, \t_\m,m_\m)$ as the fundamental fields of the gauge theory formulation of Newton-Cartan gravity. Finally, one imposes two further constraints on the curvatures of $\o_\m{}^{ab}$ and $\o_\m{}^a$ and recover the Poisson equation of Newtonian gravity and the geodesic equation for a massive particle from a gauge theory viewpoint.

In three space-time dimensions, the Einstein gravity can be established as a Chern-Simons gauge theory in which case the necessary curvature constraints arise as an equation of motion \cite{3dWitten}. This line of thinking encourages one to build up a similar three-dimensional framework for the Newton-Cartan gravity. Indeed in \cite{EBG1} this problem was addressed and it was shown that the three-dimensional Bargmann algebra does not admit a non-degenerate invariant bi-linear form to construct a Chern-Simons theory of gravity but an extension of the Bargmann algebra with a second central charge $S$ is necessary. The resulting algebra is called the extended Bargmann algebra and the corresponding Chern-Simons model is called the extended Bargmann gravity (EBG) \cite{EBG1}. The EBG is in many ways different than Newton-Cartan gravity. First of all, while the Newton-Cartan gravity dictates that only the purely time-like component of the Ricci tensor is non-zero (\ref{NewtonCartan}), EBG allows all components of the Ricci tensor to be non-zero in the presence of matter coupling \cite{Bergshoeff EBG}. Second, the extended Bargmann algebra further allows a non-relativistic conformal extension and a corresponding action principle, called the extended Schr\"odinger gravity \cite{Hartong EBG}. This extension is noteworthy as it corresponds to a conformal non-projectable Horava-Lifshitz gravity \cite{Hartong EBG} through a map between the Newton-Cartan geometry and Horava-Lifshitz gravity \cite{Horava}. Finally, the extended Bargmann algebra admits a supersymmetric extension and a corresponding on-shell Chern-Simons supergravity action \cite{Bergshoeff EBG}. The existence of a supersymmetric action is particularly important as it signals the possibility of an off-shell action for the extended Bargmann supergravity which can serve as a starting point to construct non-relativistic field theories on curved backgrounds by means of localization \cite{Local1,Local2}. 

The on-shell $\cN=2$ extended Bargmann supergravity has the following field content \cite{Bergshoeff EBG}
\bea
\{e_\m{}^a, \t_\m, m_\m, s_\mu, \p_\m^+, \p_\m^-, \rho_\m \} \,,
\eea
where the fermionic fields $\p^\pm_\m$ and $\rho_\m$ are Majorana fermions that are associated with the fermionic generators $Q^\pm$ and $R$, respectively. Here, we used the $\cN=2$ terminology as the on-shell EBG can be derived by a Lie algebra expansion of the three-dimensional $\cN=2$ Poincar\'e supergravity \cite{LAE5}. The construction of an off-shell extended Bargmann supergravity requires the addition of auxiliary fields and the application of localization techniques requires the knowledge of off-shell matter multiplet actions of extended Bargmann superalgebra. In the relativistic context, this is most straightforwardly achieved by applying superconformal tensor calculus \cite{SCTC1,SCTC2,SCTC3,SCTC4}, where one first constructs superconformal models then gauge fix the redundant conformal symmetries to obtain a super-Poincar\'e invariant theory. In the case of non-relativistic gravity, a methodology for a conformal tensor calculus was established for the bosonic and supersymmetric conformal extension of the Galilei algebra, known as the Schr\"odinger algebra, in \cite{AfsharBergshoeff} and \cite{SuperSch}, respectively.

In this paper, our purpose is to take a pioneering step towards an extended Schr\"odinger tensor calculus by establishing the Schr\"odinger extension of the extended Bargmann algebra and the corresponding extended Schr\"odinger supergravity. Due to the map between the Newton-Cartan geometry and Horava-Lifshitz gravity, this result also corresponds to a superconformal non-projectable Horava-Lifshitz gravity. To achieve the extended Schr\"odinger superalgebra, we first remind the reader in Section \ref{Section2} about the fundamentals of the extended Bargmann algebra and its cosmological extension, known as the extended Newton-Hooke algebra. In Section \ref{Section3}, we construct the cosmological extension of the extended Bargmann superalgebra. The cosmological extension of the algebra allows us to construct the extended Newton-Hooke and the extended exotic $\cN=(2,0)$ Bargmann supergravity. Here, $\cN=(2,0)$ terminology refers to the fact that the non-relativistic algebra can be understood as the Lie algebra expansion of the three-dimensional $\cN=(2,0)$ AdS algebra. In Section \ref{Section4} we start adding dilatation and non-relativistic special conformal generators to the extended Bargmann algebra. In doing so, we first add the dilatations $D$ and a central charge $Y$. Perhaps unexpectedly, we found that the superalgebra can be closed by adding two $R-$symmetry generators and three fermionic generators $Q^\pm$ and $R$. The presence of the central charge $Y$ fixes the so-called dynamical critical exponent $z$ to $z=2$ and we establish $z=2$ extended Lifshitz supergravity. In the final step, we add the non-relativistic special conformal generator $K$ and a second central charge $Z$. The supersymmetric extension of the extended Schr\"odinger algebra requires two additional fermionic generators, which we call $F^\pm$. With the addition of the new fermionic generators, we close the superalgebra and construct the Chern-Simons action for the extended Schr\"odinger gravity.

\section{Extended Newton-Hooke Algebra and Chern-Simons Actions} \label{Section2}

In this first section, we briefly review the extended Newton-Hooke gravity and extended exotic Bargmann gravity as well as the underlying symmetry algebra that is the extended Newton-Hooke algebra. The Bargmann algebra consists of the generators of spatial rotations $J$, spatial translations $P_a$, Galilean boosts $G_a$, time translations $H$ and a central charge $M$. The extended Bargmann algebra extends the Bargmann algebra with a second central charge $S$ \cite{EBG1,Bergshoeff EBG,Hartong EBG}. The non-vanishing commutation relations for the extended Bargmann algebra are given by
\begin{align}
\left[ H, G_a \right]  &= - \e_{ab} P^b\,,          &  \left[ J, P_a \right] &= - \e_{ab} P^b\,,          &  \left[ J, G_a \right] &=- \e_{ab} G^b\,, \nonumber\\
\left[ G_{a}, P_b \right]  &= \e_{ab} M\,,        &  \left[ G_a, G_{b} \right]  &= \e_{ab} S \,,
\label{EBA}
\end{align}
Here, we use the conventions of \cite{Bergshoeff EBG}. For reader’s convenience, we present a map between different conventions in Appendix \ref{AppB}. The cosmological extension of the extended Bargmann algebra requires two extra commutators 
\bea
&&   \left[ H, P_a \right] = - \frac{1}{\ell^2} \e_{ab} G^b \,,\quad \left[ P_a, P_a \right] = \frac{1}{\ell^2} \e_{ab} S \,.
\label{CEBA}
\eea
Note that we set $\Lambda = \frac{1}{\ell^2}$ for future purposes, however the opposite sign is equally well to construct the cosmological bosonic models.  The construction of actions for this symmetry algebra is based on the Chern-Simons action 
\bea
S = \frac{k}{4\pi} \int \rm{STr} \left(A \wedge dA + \frac23 A \wedge A \wedge A \right) \,,
\label{CS}
\eea
where $k$ is the Chern-Simons coupling constant, the gauge field A represents a Lie-algebra-valued one form and ``STr" represents the supertrace using the non-degenerate invariant bilinear form on the Lie algebra, which turns into the standard trace for the purely bosonic models. For the extended Bargmann algebra, the gauge field $A_\m$ is given by
\bea
A_\m = \t_\m H + e_\m{}^a P_a + \o J + \o_\m{}^a G_a + m_\m M + s_\m S \,.
\eea
The transformation rules for these gauge fields can be found by the standard rule
\bea
\d A_\m^A = \partial_\m \e^A + f_{BC}{}^A \e^C A_\m^B \,,
\label{StandardRule}
\eea
where $\e^A$ is the relevant gauge parameter and $f_{BC}{}^A$ are the structure constants. The covariant curvatures are given by
\bea
R_{\m\n}(H) &=& 2 \partial_{[\m} \t_{\n]}   \,,\nn\\ 
R_{\m\n}{}^a (P) &=&  2 \partial_{[\m} e_{\n]}{}^a + 2 \e^{ab}\,  \o_{[\m} e_{\n]b} - 2 \e^{ab}\,  \o_{ [\m b} \t_{\n]} \,,\nn\\ 
R_{\m\n} (J) &=& 2 \partial_{[\m} \o_{\n]} \,,\nn\\
R_{\m\n}{}^a (G) &=& 2 \partial_{[\m} \o_{\n]}{}^a + 2 \e^{ab} \o_{[\m} \o_{\n]b} + \frac{2}{\ell^2} \e^{ab} \t_{[\m} e_{\n]b}  \,,\nn\\ 
R_{\m\n}(M) &=& 2 \partial_{[\m} m_{\n]} + 2 \e_{ab} \o_{[\m}{}^a e_{\n]}{}^b   \,,\nn\\ 
R_{\m\n}(S) &=& 2 \partial_{[\m} s_{\n]} +  \e_{ab} \o_{[\m}{}^a \o_{\n]}{}^b + \frac{1}{\ell^2}  \e_{ab} e_{[\m}{}^a e_{\n]}{}^b  \,.
\label{Curvatures1}
\eea
Note that for $\ell \rightarrow \infty$, these curvatures become the curvatures of the extended Bargmann algebra.

As mentioned, the construction of a Chern-Simons action requires non-degenerate invariant bilinear forms which can simply be found by constructing a Casimir operator $C$. In the case of extended Newton-Hooke algebra the following bilinears are of interest
\begin{itemize}
\item {$(P_a,G_b) = \delta_{ab}\,, \quad (H,S) =-1\,, \quad (J,M) = -1$}
\item{$(P_a,P_b) =  \frac{1}{\ell} \delta_{ab} \,, \quad (G_a,G_b) = \ell  \delta_{ab}\,, \quad (H,M) =- \frac{1}{\ell} \,,\quad  (J,S) = -\ell\,, $}
\end{itemize}
In the first case, the cosmological constant does not appear in the bilinears hence it can also be used to construct the extended Bargmann gravity without a cosmological constant. On the other hand, the second set only exists in the presence of a cosmological constant. A more detailed discussion on the invariant bilinear forms, including the degenerate ones in context of extended Bargmann algebra, can be found in \cite{Hartong EBG}. 

\subsection{Extended Newton-Hooke Gravity}

For the construction of the extended Newton-Hooke gravity we utilize the first set of invariant bilinear forms
\bea
(P_a, G_b) = \d_{ab} \,, \quad (H,S) =  -1 \,, \quad (J,M) = -1\,,
\label{Bilinear1}
\eea
and the structure constants of the extended Newton-Hooke algebra (\ref{EBA}) and (\ref{CEBA}). In this case, the Chern-Simons action (\ref{CS}) read 
\bea
S &=&\frac{k}{4 \pi} \int d^3x\, \e^{\m\n\r}  \left( e_\m{}^a R_{\n\r a} (G) - \t_\m R_{\n\r}(S) - m_\m R_{\n\r}(J) + \frac{2}{\ell^2} \e_{ab} \t_\m e_\n{}^a e_\r{}^b \right) \,, \quad 
\label{CEBG} 
\eea
where the curvatures are as defined in (\ref{Curvatures1}). For $\ell \rightarrow \infty$, this action coincides with the extended Bargmann gravity of \cite{EBG1,Bergshoeff EBG,Hartong EBG}. Interestingly,  the action for the extended Newton-Hooke gravity can also be obtained by a contraction procedure, mimicking the Inonu-Wigner contraction  \cite{Bergshoeff EBG}\footnote{Extended Bargmann algebra can also be obtained by the In\"on\"u-Wigner contraction of the Poincar\'e $\otimes$ $U(1)^2$ algebra, see \cite{MaxwellContraction,Color} }. In order to do so, one needs to consider the cosmological Einstein-Hilbert action plus a Chern-Simons action for two abelian gauge fields $Z_{1\m}$ and $Z_{2\m}$ with an off-diagonal coupling \cite{Bergshoeff EBG}
\bea
S &=&   \frac{k \o}{4 \pi }\int  d^3 x \  \epsilon^{\m \n \r} \left( \eta_{AB}  E_\m{}^A R_ {\n\r}{}^B(\O) + \frac{1}{3L^2}\   \e_{ABC}\ E_\m{}^A   E_\n{}^B E_\r{}^C + 2 Z_{1 \m} \partial_\n Z_{2 \r}   \right)\,, \label{EHV}
\eea
where
\bea
R_{\m\n}{}^A (\O) &=& 2 \partial_{ [ \mu } \Omega_{\n]}{}^A  + \e^{ABC} \O_{\m B} \O_{\n C} \,.
\eea
In terms of the non-relativistic fields and parameters, the relativistic fields and parameters are  \cite{Bergshoeff EBG}
\bea
&& E_\m{}^a = e_\m{}^a  \,, \quad E_\m{}^0  = \o  \t_\m + \frac{1}{2\o} m_\m  \,, \quad   \O_\m{}^0= \o_\m + \frac{1}{2 \o ^2 } s_\m \,,\quad  L = \o \ell \,, \nn \\
&&  \O_\m{}^a  = \frac{1}{\o}\o_\m{}^a\,, \quad Z_{1 \m} = \o \t_\m - \frac{1}{2 \o } m_\m \,, \quad Z_{2\m} = \o_\m - \frac{1}{2 \o^2}s_\m \,.
\label{contraction1}
\eea
The off-diagonal coupling of the vector fields are essential for removing the $\o^2$ divergencies. In the $\omega \rightarrow \infty$ limit, the action (\ref{EHV}) is precisely recover the extended Newton-Hooke gravity action (\ref{CEBG}). 

\subsection{Extended Exotic Bargmann Gravity}

In three dimensions there is an alternative action that gives rise to the Einstein's equations in the presence of a cosmological constant, known as the exotic Einstein-Hilbert action \cite{3dWitten,TownsendExotic}
\bea
S =  \frac{kL \o  }{2 \pi }\int  d^3 x  \  \epsilon^{\m \n \r} \left(\O_\m{}^A \partial_ \n \O_{\r A}  + \frac{1}{3} \e_{ABC} \O_\m{}^A \O_\n{}^B \O_\r{}^C  +   \frac{1}{2 L^2} E_\m{}^A  R_{\n\r A}(P)  \right) \,.
\label{ExoticEinstein}
\eea 
This indicates that there might be an extended ``exotic" Bargmann gravity following the contraction procedure introduced in (\ref{contraction1}). As the exotic theory has diagonal coupling amongst the gauge fields of the Poincar\'e algebra, the vector fields that are neccessary to cancel out the $\o^2$ divergences should also appear with a diagonal coupling. Thus, we consider the exotic Einstein-Hilbert action plus a diagonal Chern-Simon action for two abelian gauge fields
\bea
S &=& \frac{kL\o}{2 \pi }\int  d^3 x  \  \epsilon^{\m \n \r} \Big(  \O_\m{}^A \partial_ \n \O_{\r A}  + \frac{1}{3} \e_{ABC} \O_\m{}^A \O_\n{}^B \O_\r{}^C  +   \frac{1}{2 L^2} E_\m{}^A  R_{\n\r A}(P) \nn\\
&&  \quad   + \frac{1}{L^2} Z_{1\m} \partial_\n Z_{1\r}  + Z_{2\m} \partial_\n Z_{2\r} \Big) \,,
\label{ExoticZZ}
\eea
where 
\bea
R_{\m\n}{}^A (P) &=& 2 \partial_{[\m} E_{\n]}{}^A + 2 \e^{ABC} \O_{[\m B} E_{\n] C} \,.
\eea
Using the expressions in (\ref{contraction1}) in the action (\ref{ExoticZZ}), we find that the extended exotic Bargmann gravity is given by
\bea
S &=& \frac{k\ell}{4 \pi} \int d^3x\, \e^{\m\n\r}  \Big( \o_\m{}^a R_{\n\r a} (G) -  2 s_\m R_{\n\r}(J)  + \frac{1}{\ell^2}  \left( e_\m{}^a R_{\n\r a} (P) - 2 m_\m R_{\n\r} (H) \right)  \Big)\,. \qquad \quad 
\label{EEBG}
\eea
At this stage, it is worthwhile to discuss the ``exotic" terminology in the context of relativistic and non-relativistic gravity. In the relativistic context, the fundamental fields $E^a$ and $\Omega^{ab}$ are parity even, thus the dual spin connection, $\O^a$ is parity odd. As a result the Einstein-Hilbert action (\ref{EHV}) is parity even while the exotic Einstein-Hilbert action (\ref{ExoticEinstein}) is parity odd. Here, the exotic terminology emphasizes the fact that although the exotic Einstein-Hilbert action is of odd-parity, it give rise to parity even equations of motion. In the non-relativistic context, we have the following parity assignment for the gauge fields
\bea
\text{Even:} \quad   \{ \t_\m \,, \quad e_\m{}^a\,, \quad m_\m \} \,, \qquad \qquad \text{Odd:}  \quad  \{ \o_\m \,, \quad \o_\m{}^a\,, \quad s_\m  \} \nn \,.
\eea
In this case, the extended Newton-Hooke gravity is parity even while the extended exotic Bargmann gravity is parity odd. As both of these theories have parity even equations of motion, we use the ``exotic" terminology for the action (\ref{EEBG}). It is also possible to obtain the extended exotic Bargmann gravity by using the second set of bilinear forms \cite{Hartong EBG}
\bea
(P_a, P_b) =  \frac{1}{\ell^2} \d_{ab} \,, \quad (G_a, G_b) = \d_{ab} \,, \quad (H,M) = - \frac{1}{\ell^2}\,,\quad (J,S) = -1 \,,
\eea
and tracing the Chern-Simons action. The resulting action precisely coincide with the extended exotic Bargmann gravity (\ref{EEBG}). 

Although the field equations for the extended Newton-Hooke gravity (\ref{CEBG}) and the  extended exotic Bargmann gravity (\ref{ExoticEBG}) are the same, these two models differ dramatically when we consider the matter couplings. In particular, let us first take a closer look at the $s_\m, \o_\m{}^a$ and $\o_\m$ equations for the  extended Newton-Hooke gravity
\bea
R_{\m\n}(H) = 0 \,, \qquad  R_{\m\n}{}^a (P) = 0 \,, \qquad R_{\m\n}(M) = 0 \,.
\eea
The first equation implies that this model is defined on non-relativistic space-times with torsionless Newton-Cartan geometry while the last two equations give rise to composite expressions for $\o_\m$ and $\o_\m{}^a$. The torsionless condition is imposed employing $s_\m$ equation, which suggests that one needs to find a matter content that transforms non-trivially under $S$-transformations to enable torsional non-relativistic background geometries. On the other hand, the same torsion-free constraint is imposed by the $m_\m$ equation in the case of extended exotic Bargmann gravity. Accordingly one needs to find a  matter content that transforms non-trivially under $M$-transformations to include torsion. As opposed to the extended Bargmann gravity, this can be simply achieved, for instance, by coupling a complex scalar field to extended exotic Bargmann gravity. Consequently, extended exotic Bargmann gravity provides a more useful setup to study three-dimensional torsional non-relativistic geometries. 

\section{Extended Newton-Hooke and Exotic Bargmann Supergravity} \label{Section3}

The $\cN=(2,0)$ supersymmetric extension of the extended Newton-Hooke algebra, (\ref{EBA}) and (\ref{CEBA}), requires three supercharges $Q^\pm_\a$ and $R_\a$  $(\a = 1,2)$ that are all Majorana spinors. It furthermore requires two extra bosonic generators $U_1$ and $U_2$. Both $U_1$ and $U_2$ act non-trivially on the spinors $Q^\pm$ and $R$ in the presence of the cosmological constant and they become central when cosmological constant vanishes. The $\cN=(2,0)$ extended Newton-Hooke superalgebra has the following non-vanising $\left[B,B\right]$ and $\left[B,F\right]$ commutators
\begin{align}
\left[ H, G_a \right]  &= - \e_{ab} P^b\,,          &  \left[ J, P_a \right] &= - \e_{ab} P^b\,,          &  \left[ J, G_a \right] &=- \e_{ab} G^b\,, \nonumber\\
\left[ G_{a}, P_b \right]  &= \e_{ab} M\,,        &  \left[ G_a, G_{b} \right]  &= \e_{ab} S \,, &    \left[ P_a, P_{b} \right]  &= \frac{1}{\ell^2} \e_{ab} S   \,, \nn\\
\left[ H, P_a \right]  &= - \frac{1}{\ell^2} \e_{ab} G^b\,, & [J, Q^\pm ] & =  - \frac{1}{2} \g_0 Q^\pm  \,,  & [J, R ] &= - \frac{1}{2} \g_0 R \,,\nn\\
[S, Q^+ ] & = - \frac{1}{2} \g_0 R \,, & [G_a, Q^+] &= - \frac{1}{2} \g_a Q^- \,,& [G_a, Q^-] &= - \frac{1}{2} \g_a R \,,\nn\\
[P_a, Q^+] &= - \frac{1}{2\ell} \g_a Q^-\,, & [P_a, Q^-] &= - \frac{1}{2\ell} \g_a R \,, & [H, Q^\pm] &= - \frac{1}{2\ell} \g_0 Q^\pm \,,\nn\\
[H, R] &= - \frac{1}{2\ell} \g_0 R \,, & [M, Q^+ ] &= - \frac{1}{2\ell} \g_0 R \,, &[U_1, Q^\pm] &= \mp \frac{1}{\ell}  \g_0 Q^\pm \,,\nn\\
[U_1, R] &= - \frac{1}{\ell}  \g_0 R\,, & [U_2, Q^+ ] &= - \frac{1}{\ell}  \g_0 R \,.
\label{CosmoEBG1}
\end{align}
while the non-vanishing $\{F, F\}$ anti-commutators are given by
\bea
\{ Q^+_\a ,  Q^+_\b  \} &=& (\g_0 C^{-1})_{\a \b} H  +   \frac{1}{\ell}  (\g_0 C^{-1})_{\a \b} J  -   (\g_0 C^{-1})_{\a \b} U_1  \,,\nonumber\\
\{ Q^+_\a ,  Q^-_\b  \}  &=& -(\g_a C^{-1})_{\a \b} P^{a} -  \frac{1}{\ell}   (\g_a C^{-1})_{\a \b} G^{a}\,,\nonumber\\
\{ Q^+_\a ,  R_\b  \} &=& (\g_0 C^{-1})_{\a \b} M  +  \frac{1}{\ell} (\g_0 C^{-1})_{\a \b} S  -   (\g_0 C^{-1})_{\a \b} U_2\,,\nn \\
\{ Q^-_\a ,  Q^-_\b  \} &=& (\g_0 C^{-1})_{\a \b} M  +   \frac{1}{\ell}  (\g_0 C^{-1})_{\a \b} S  +  (\g_0 C^{-1})_{\a \b} U_2  \,.
\label{CosmoEBG2}
\eea
The appearence of the extra bosonic generators $U_{1,2}$ can be understood as  the Lie algebra expansion of the $R$-symmetry generator of the $\cN=(2,0)$ AdS superalgebra, which we defer the details to the Appendix \ref{AppA}. Here, our splitting of the indices of gamma matrices into its temporal and the spatial components as well as our treatment to gamma-matrices in general follow from \cite{SezBerg}. As in the bosonic case, the extended Newton-Hooke algebra admit two distinct non-degenerate bilinear forms. In the first case, it is given by  
\bea
&& (P_a,G_b) = \delta_{ab}\,, \quad (H,S) =-1\,, \quad (J,M) = -1\,, \quad (U_1, U_2) = \frac{2}{\ell}\,,\nn\\
&&  (Q^+_\alpha,R_\beta)= 2 \left( C^{-1}\right)_{\a \b}\,,\quad (Q^-_\alpha,Q^-_\beta)= 2 \left( C^{-1}\right)_{\a \b} \,,
\label{MetricCEBG1}
\eea
Taking the gauge field to be
\bea
A_\m &=& \t_\m H + e_\m{}^a P_a + \o_\m J + \o_\m{}^a G_a + m_\m M + s_\m S +r_{1\m} U_1 + r_{2\m} U_2 \nn\\
&& + \bar\p_\m^+ Q^+ + \bar\p_\m^- Q^- + \bar\rho_\m R \,,
\label{GaugeField2}
\eea
we can construct a Chern-Simons action by using the invariant bilinear form (\ref{MetricCEBG1}). The resulting theory is the extended Newton-Hooke supergravity. This action is given by
\bea
S &=&\frac{k}{4 \pi} \int d^3x\, \ve^{\m\n\r}  \Big( e_\m{}^a R_{\n\r a} (G) - \t_\m R_{\n\r} (S) - m_\m R_{\n\r}(J) + \frac{2}{\ell^2}  \e_{ab} \t_\m e_\n{}^a e_\r{}^b \nn \\
&&  + \frac{4}{\ell} r_{1\m} \partial_{\n} r_{2\r} +  \bar\p_\m^+ R_{\n\r} (R) +   \bar\r_\m R_{\n\r} (Q^+) +  \bar\p_\m^- R_{\n\r} (Q^-) \Big) \,.
\label{CosmoEBG}
\eea
The bosonic part of this action, excluding the $R$-symmetry part, matches with the extended Newton-Hooke action given in \cite{Hartong EBG}. Here, the bosonic curvatures are as given in (\ref{Curvatures1}) while the fermionic curvatures are given as
\bea
R_{\m\n} (Q^+) &=& 2 \partial_{[\m} \p_{\n]}^+ +  \o_{[\m} \g_0 \p_{\n]}^+   + \frac{1}{\ell}  \t_{[\m} \g_0 \p_{\n]}^+ +   \frac{2}{\ell} r_{1[\m} \g_0 \p_{\n]}^+  \,,\nn\\ 
R_{\m\n} (Q^-) &=& 2 \partial_{[\m} \p_{\n]}^- +  \o_{[\m} \g_0 \p_{\n]}^-  + \o_{[\m}{}^a \g_a \p_{\n]}^+    + \frac{1}{\ell} \t_{[\m} \g_0 \p_{\n]}^-  + \frac{1}{\ell} e_{[\m}{}^a \g_a \p_{\n]}^+ -  \frac{2}{\ell}  r_{1[\m} \g_0 \p_{\n]}^-  \,,\nn\\
R_{\m\n} (R) &=& 2 \partial_{[\m} \rho_{\n]} +  \o_{[\m} \g_0 \rho_{\n]}  + \o_{[\m}{}^a \g_a \p_{\n]}^- +  s_{[\m} \g_0 \p_{\n]}^+  + \frac{1}{\ell} \t_{[\m} \g_0 \rho_{\n]}  + \frac{1}{\ell} m_{[\m} \g_0 \p_{\n]}^+   \nn\\
&&  +  \frac{2}{\ell} r_{1[\m} \g_0 \rho_{\n]}  +   \frac{1}{\ell} e_{[\m}{}^a \g_a \p_{\n]}^-  +  \frac{2}{\ell} r_{2[\m} \g_0 \p_{\n]}^+  \,.
\eea
In the presence of a cosmological constant, there is a second non-degerate invariant bilinear form that is given by
\bea
&&(P_a,P_b) =  \frac{1}{\ell} \delta_{ab} \,, \quad (G_a,G_b) = \ell  \delta_{ab}\,, \quad (H,M) =- \frac{1}{\ell} \,,\quad  (J,S) = -\ell\,, \nn \\
&&(U_1,U_2) = \frac{2}{\ell}\,, \quad (Q^+_\alpha,R_\beta)= 2 \left( C^{-1}\right)_{\a \b}\,,\quad (Q^-_\alpha,Q^-_\beta)= 2 \left( C^{-1}\right)_{\a \b} \,,
\eea
In this case, we obtain the ``exotic" extended Bargmann supergravity
\bea
S &=& \frac{k \ell}{4\pi} \int d^3x\, \e^{\m\n\r}  \Big(\o_\m{}^a R_{\n\r a} (G) -  2 s_\m R_{\n\r}(J) + \frac{1}{\ell^2}  e_\m{}^a R_{\n\r a} (P) -  \frac{2}{\ell^2} \t_\m R_{\n\r} (M)  \nn \\
&&  \quad + \frac{8}{\ell^2} r_{2\m} \partial_\n r_{1\r}  +  \frac{2}{\ell}  \bar\p_\m^+ R_{\n\r} (R)  +  \frac{2}{\ell}  \bar\r_\m R_{\n\r} (Q^+) + \frac{2}{\ell}   \bar\p_\m^- R_{\n\r} (Q^-)  \Big) \,.
\label{ExoticEBG}
\eea
The bosonic part of this action, excluding the $R$-symmetry part, matches with the extended exotic Bargmann action given in \cite{Hartong EBG}. The extended Newton-Hooke (\ref{CosmoEBG}) and the exotic extended Bargmann gravity (\ref{ExoticEBG}) are invraiant under the following supersymmetry transformation rules
\bea
\d \t_\m &=&  - \bar{\e}^+ \g_0 \p^+_\mu \,,\nn\\
\d e_\m{}^a &=&    \bar{\e}^+ \g^{a} \p_\m^- +  \bar{\e}^- \g^{a} \p_\m^+ \,,\nn\\
\d \o_\m{}^a &=&  \frac{1}{\ell}(\bar{\e}^+ \g^{a} \p_\m^- +  \bar{\e}^- \g^{a} \p_\m^+) \,,\nn\\
\d \o_\m &=& - \frac{1}{\ell} \bar{\e}^+ \g_0 \p^+_\mu \,,\nn\\
\d m_\m &=& - \bar{\e}^- \g_0 \p_\m^- -  \bar{\e}^+ \g_0 \rho_\m -  \bar{\eta} \g_0 \p_\m^+  \,,\nn\\
\d s_\m  &=& - \frac{1}{\ell} \left( \bar{\e}^- \g_0 \p_\m^- + \bar{\e}^+ \g_0 \rho_\m + \bar{\eta} \g_0 \p_\m^+\right) \,,\nn \\
\d r_{1\m}  &=&   \bar{\e}^+ \g_0 \p^+_\mu \,, \nn \\
\d r_{2\m}  &=&  -  \bar{\e}^- \g_0 \p_\m^- +  \bar{\e}^ + \g_0 \rho_\m +  \bar{\eta} \g_0 \p_\m^+ \,, \nn \\
\d \p_\m^+ &=& \partial_\m \e^+ + \frac12 \o_\m \g_0 \e^+  +  \frac1{2\ell} \t_\m \g_0 \e^+ + \frac{1}{\ell} r_{1\m} \g_0 \e^+ \,,\nn\\
\d \p_\m^- &=&  \partial_\m \e^- + \frac12 \o_\m \g_0 \e^- + \frac12 \o_\m{}^a \g_a \e^+ +  \frac{1}{2\ell} \t_\m \g_0 \e^-  +  \frac{1}{2\ell}  e_\m^a \g_a \e^+ -  \frac{1}{\ell}  r_{1\m} \g_0 \e^- \,,\nn\\
\d \r_\m &=&  \partial_\m \eta + \frac12 \o_\m \g_0 \eta + \frac12 \o_\m{}^a \g_a \e^- + \frac12 s_\m \g_0 \e^+ +   \frac{1}{2\ell}  \t_\m \g_0 \eta + \frac{1}{2\ell}  m_\m \g_0 \e^+  \nn\\
&& +  \frac{1}{2\ell}  e_\m^a \g_a \e^- +   \frac{1}{\ell}  r_{1\m} \g_0 \eta  +  \frac{1}{\ell}  r_{2\m} \g_0 \e^+  \,,
\label{CosmoGaugeEBG} 
\eea
where $\e^\pm$ and $\eta$ are the the parameters of the local $Q^\pm$ and $R$ transformations, respectively. In the supersymmetric extension of the extended Newton-Hooke gravity (or extended exotic Bargmann gravity), the curvature constraints that are given by the $s_\m, \o_\m{}^a, \o_\m$ (or $m_\m, e_\m{}^a, \t_\m$) are given by
\bea
R_{\m\n}(H) = 0 \,, \qquad R_{\m\n}{}^a (P) = 0 \,, \qquad R_{\m\n}(M) = 0 \,,
\eea
where 
\bea
R_{\m\n}(H) &=& 2 \partial_{[\m} \t_{\n]}  + \bar\p_\m \g_0 \p_\n   \,,\nn\\ 
R_{\m\n}{}^a (P) &=&  2 \partial_{[\m} e_{\n]}{}^a + 2 \e^{ab}\,  \o_{[\m} e_{\n]b} - 2 \e^{ab}\,  \o_{ [\m b} \t_{\n]} - 2 \bar\p_{[\m}^+ \g^a \p_{\n]}^-  \,,\nn\\ 
R_{\m\n}(M) &=& 2 \partial_{[\m} m_{\n]} + 2 \e_{ab} \o_{[\m}{}^a e_{\n]}{}^b + 2 \bar\p_{[\m}^+ \g_0 \r_{\n]} + \bar\p_\m^- \g_0 \p_\n^-   \,.
\eea
In this case, first equation imply a fermionic temporal torsion while the last two equations gives rise to composite expressions for $\o_\m$ and $\o_\m{}^a$ that includes a fermionic part. The back substitution of these composite expressions into the action (\ref{CEBG}) (or (\ref{ExoticEBG})) lead to quartic fermionic terms in the action.

\section{Extended Lifshitz and Schr\"odinger Supergravity} \label{Section4}

In this section, we take our final step towards the completion of the realm of three-dimensional $\cN=2$ extended Bargmann supergravity. The missing pieces are models of supergravity with more symmetries such as dilatations and non-relativistic special conformal symmetry.  In the first subsection, we show that it is possible to extend the extended Bargmann algebra with dilatations and a central charge, $Y$, giving rise to the extended Lifshitz superalgebra and consequently the extended Lifshitz supergravity. We then proceed with the full Schr\"odinger superalgebra and construct a Chern-Simons action as an extended Schr\"odinger supergravity. 

\subsection{Extended Lifshitz Supergravity}

The inclusion of dilatations ($D$) in the extended Bargmann algebra (\ref{EBA}) requires an additional central charge ($Y$)  in order to have a non-degenerate metric. The non-vanishing commutation relations for this algebra are given by
\begin{align}
\left[ H, G_a \right]  &= - \e_{ab} P^b\,,          &  \left[ J, P_a \right] &= - \e_{ab} P^b\,,          &  \left[ J, G_a \right] &=- \e_{ab} G^b\,, \nonumber\\
\left[ P_{a}, G_b \right]  &= \e_{ab} M + \d_{ab} Y \,,        &  \left[ G_a, G_{b} \right]  &=  \e_{ab} S \,, & \left[ D, S \right]  &=2 S \,,  \nn\\
\left[ D, G_a \right]  &= G_{a} \,,          &  \left[ D, P_a \right] &= -  P_{a} \,,          &  \left[ D, H \right] &=- 2 H\,, \nonumber\\ 
\left[ H, S \right]  &= - 2 Y  \,,         
\label{LifBoson}
\end{align}
We refer to this algebra as the extended extended Lifshitz algebra. Note here that the Lifshitz symmetry usually allow the scaling dimension of the temporal translations $H$ to be arbitrary, which is referred to as the dynamical critical exponent $z$. However, the inclusion of the central charge $Y$ fixes the scaling dimension of $H$ to $z=2$. This algebra allows the following non-degenerate invariant bi-linear form
\bea
(P_{a}, G_b) = \d_{ab} \,, \quad (S,H) = -1 \,, \quad (M, J) = -1\,, \quad (Y,D) = -1 \,. \label{metricbosonicLifshitz}
\eea
Taking the gauge field to be
\bea
A_\m &=& \t_\m H + e_\m{}^a P_a + \o_\m J + \o_\m{}^a G_a + m_\m M + s_\m S + b_\mu D + y_\m Y \,,
\label{GFLifBos}
\eea
 we can use to construct a Chern-Simons action for the extended Lifshitz algebra
\bea
S &=&\frac{k}{4 \pi} \int d^3x\, \ve^{\m\n\r}  \Big( \o_\m{}^a  R_{\n\r a} (P) - 2 s_\m  R_{\n\r} (H) - 2 m_\m  R_{\n\r}(J) \nn\\
&& \quad + e_\m{}^a  R_{\n\r a} (G)   - 2 y_\m  R_{\n\r} (D)  \Big) \,.
\label{BosonicLifshitzAction}
\eea
Here, the curvatures of the gauge fields are given by
\bea
R_{\n\r}{}^a (P) &=& 2 \partial_{[\m} e_{\n]}{}^a + 2 \e^{ab}\,  \o_{[\m} e_{\n]b} - 2 \e^{ab}\,  \o_{ [\m b} \t_{\n]}  - 2 b_{[\m} e_{\n]}^a \,,\nn\\
R_{\n\r}{}^a (G) &=&   2 \partial_{[\m} \o_{\n]}{}^a + 2  \epsilon{}^{ab} \o_{[\m}{} \o_{\n] b} + 2 b_{[\m} \o_{\n]}^a \,\nn\\
R_{\n\r} (H) &=&  2 \partial_{[\m} \t_{\n]}  - 4 b_{[\m} \t_{\n]}\,,\nn\\
R_{\n\r}(J) &=&  2 \partial_{[\m} \o_{\n]} \,,\nn\\
R_{\n\r} (D) &=& 2 \partial_{[\m} b_{\n]}   \,. \label{LifshitzBosonicCurvatures}
\eea
As the non-relativistic special conformal symmetry is absent, the supersymmetric extension of the extended Lifshitz algebra does not require a special supersymmetry generator but the three fermionic generators $Q^\pm$ and $R$ that we introduced in the previous sections are sufficient. The inclusion of the generators $U_{1,2}$ is still necessary, but while $U_1$ acts non-trivially on the fermionic generators, $U_2$ is a central charge in the extended Lifshitz superalgebra. With the assistance of the computer algebra program \textit{Cadabra} \cite{Cadabra1,Cadabra2}, we found that the non-vanishing $\left[B,B\right]$ commutators of the extended Lifshitz superalgebra are given by (\ref{LifBoson}) while the non-vanishing $\left[B,F\right]$ commutators are
\begin{align}
[J, Q^\pm ] &= - \frac{1}{2} \g_0 Q^\pm  \,, & [J, R] &= - \frac{1}{2} \g_0 R  \,, &[G_a, Q^+  ] &= - \frac{1}{2} \g_a Q^-  \,,\nn\\
[G_a, Q^-  ] &= - \frac{1}{2} \g_a R \,, & [U_1, Q^\pm ] &= \pm  \g_0 Q^\pm  \,, &  [U_1, R ] &=    \g_0 R \,,\nn\\
[D, Q^+] &= - Q^+ \,, & [D, R] &=  R\,, & [S, Q^+ ] &= - \frac{1}{2} \g_0 R
\end{align}
Finally, the non-vanishing $\{F,F\}$ anti-commutators are given by
 \bea
 \{ Q^+_\a ,  Q^+_\b  \} &=& (\g_0 C^{-1})_{\a \b} H \,,\nonumber\\
\{ Q^+_\a ,  Q^-_\b  \}  &=& -(\g_a C^{-1})_{\a \b} P^{a} \,,\nonumber\\
 \{ Q^-_\a ,  Q^-_\b  \}  &=& (\g_0 C^{-1})_{\a \b} M  - (\g_0 C^{-1})_{\a \b} U_2\,, \nn \\
\{ Q^+_\a ,  R_\b  \}  &=& (\g_0 C^{-1})_{\a \b} M + 2 C_{\alpha \beta} Y + (\g_0 C^{-1})_{\a \b} U_2 \,. 
\label{FermionicLifshitz}
 \eea
The construction of a Chern-Simons action for the extended Lifshitz supergravity can be achieved by using the following non-degenerate invariant bi-linear form 
 \bea
 && (P_{a}, G_b) = \d_{ab} \,, \quad (S,H) = -1 \,, \quad (M, J) = -1\,, \quad (Y,D) = -1 \nn\,,\\
 &&  (U_1, U_2) =  2\,,\quad  (Q^+_\alpha,R_\beta)= 2 \left( C^{-1}\right)_{\a \b}\,,\quad (Q^-_\alpha,Q^-_\beta)= 2 \left( C^{-1}\right)_{\a \b} \,.
 \label{metricfermionicLifshitz}
 \eea
 Taking the gauge field to be
 \bea
 A_\m &=& \t_\m H + e_\m{}^a P_a + \o_\m J + \o_\m{}^a G_a + m_\m M + s_\m S +r_{1\m} U_1 \nn\\
 && + r_{2\m} U_2 + b_\mu D + y_\mu Y + \bar\p_\m^+ Q^+ + \bar\p_\m^- Q^- + \bar\rho_\m R  \,,
 \label{GFSuperLif}
 \eea
we can construct a supersymmetric action
 \bea
 S &=&\frac{k}{4 \pi} \int d^3x\, \ve^{\m\n\r}  \Big( \o_\m{}^a  R_{\n\r a} (P) + e_\m{}^a  R_{\n\r a} (G)  - 2 s_\m  R_{\n\r} (H) - 2 m_\m  R_{\n\r}(J)   \nn\\
 && \quad  - 2 y_\m  R_{\n\r} (D)   +   4 r_{2\m} R_{\n\r}(U_1) +  4  \bar\r_\m R_{\n\r} (Q^+)  + 2 \bar\p_\m^- R_{\n\r} (Q^-)  \Big) \,.
 \label{FermionicLifshitzAction}
 \eea
 We refer to this action as the extended Lifshitz supergravity. The supercovariant curvatures that appears in the action are given by 
 \bea
 R_{\m\n}(H) &=& 2 \partial_{[\m} \t_{\n]}  - 4 b_{[\m} \t_{\n]} + \bar\p_\m^+ \g_0 \p_\n^+  \,,\nn\\ 
 R_{\m\n}{}^a (P) &=&  2 \partial_{[\m} e_{\n]}{}^a + 2 \e^{ab}\,  \o_{[\m} e_{\n]b} - 2 \e^{ab}\,  \o_{ [\m b} \t_{\n]}  - 2 b_{[\m} e_{\n]}^a- 2 \bar\p_{[\m}^+ \g^a \p_{\n]}^-  \,,\nn\\ 
 R_{\m\n}{}^a (G) &=& 2 \partial_{[\m} \o_{\n]}{}^a + 2  \epsilon{}^{ab} \o_{[\m}{} \o_{\n] b}  + 2 b_{[\m} \o_{\n]}^a  \,,\nn\\
 R_{\m\n} (J) &=& 2 \partial_{[\m} \o_{\n]}  \,,\nn\\
 R_{\m\n} (D) &=& 2 \partial_{[\m} b_{\n]}  \,, \nn \\
R_{\m\n}(U_1) &=& 2 \partial_{[\m} r_{1\n]}   \,,\nn\\ 
 R_{\m\n}(Q^+) &=& 2 \partial_{[\m} \p_{\n]}^+ +  \o_{[\m} \g_0 \p_{\n]}^+  - 2 b_{[\m} \p_{\n]}^+  -  2 r_{1[\m} \g_0 \p_{\n]}^+  \,,\nn\\ 
R_{\m\n}(Q^-) &=& 2 \partial_{[\m} \p_{\n]}^- +  \o_{[\m} \g_0 \p_{\n]}^-  + \o_{[\m}{}^a \g_a \p_{\n]}^+  +    2   r_{1[\m} \g_0 \p_{\n]}^-  \,,\nn\\ 
R_{\m\n}(R) &=& 2 \partial_{[\m} \rho_{\n]} +  \o_{[\m} \g_0 \rho_{\n]}  + \o_{[\m}{}^a \g_a \p_{\n]}^- +  s_{[\m} \g_0 \p_{\n]}^+  -   2 r_{1[\m} \g_0 \rho_{\n]}  + 2 b_{[\m}  \r_{\n]}  \,.
 \label{Scalesupercurvatures}
 \eea
 The extended Lifshitz supergravity is invariant under the following supersymmetry transformation rules
 \bea
 \d \t_\m &=&  - \bar{\e}^+ \g_0 \p^+_\mu \,,\nn\\
 \d e_\m{}^a &=&    \bar{\e}^+ \g^{a} \p_\m^- +  \bar{\e}^- \g^{a} \p_\m^+ \,,\nn\\
 \d m_\m &=& - \bar{\e}^- \g_0 \p_\m^- -  \bar{\e}^+ \g_0 \rho_\m -  \bar{\eta} \g_0 \p_\m^+  \,,\nn\\
 \d r_{2\m}  &=&   \bar{\e}^- \g_0 \p_\m^- -  \bar{\e}^+ \g_0 \rho_\m -  \bar{\eta} \g_0 \p_\m^+ \,,\nn\\
 \d y_\m  &=&  2 \bar{\eta} \p_\m^+ - 2 \bar{\e}^+ \r_\m  \,,\nn \\
 \d \p_\m^+ &=& \partial_\m \e^+ + \frac12 \o_\m \g_0 \e^+  - b_\m \e^+ -  r_{1\m} \g_0 \e^+ \,,\nn\\
 \d \p_\m^- &=&  \partial_\m \e^- + \frac12 \o_\m \g_0 \e^- + \frac12 \o_\m{}^a \g_a \e^+  +   r_{1\m} \g_0 \e^- \,,\nn\\
 \d \r_\m &=&  \partial_\m \eta + \frac12 \o_\m \g_0 \eta + \frac12 \o_\m{}^a \g_a \e^- + \frac12 s_\m \g_0 \e^+  -  r_{1\m} \g_0 \eta    + b_\m \eta   \,. 
 \label{GaugeLifshitzSusy} 
 \eea
\subsection{Extended Schr\"odinger Supergravity}

As mentioned, our final purpose is to establish the supersymmetric extension of the Schr\"odinger algebra. The bosonic sector of the Schr\"odinger algebra extends the Lifshitz algebra (\ref{LifBoson}) with non-relativistic special conformal transformations $(K)$ as well as a second central charge $(Z)$. The non-zero commutation relations are given by \cite{Hartong EBG}
\begin{align}
\left[ H, G_a \right]  &= - \e_{ab} P^b\,,          &  \left[ J, P_a \right] &= - \e_{ab} P^b\,,          &  \left[ J, G_a \right] &=- \e_{ab} G^b\,, \nonumber\\
\left[ P_{a}, G_b \right]  &= \e_{ab} M + \d_{ab} Y \,,        &  \left[ G_a, G_{b} \right]  &= \e_{ab} S \,, & \left[ P_{a}, P_b \right]  &= \e_{ab} Z \,,  \nn\\
\left[ D, G_a \right]  &= G_{a} \,,          &  \left[ D, P_a \right] &= -  P_{a} \,,          &  \left[ D, H \right] &=- 2 H\,, \nonumber\\
\left[ D, S \right]  &=2 S \,,        &  \left[ D,K \right]  &= 2K \,, & \left[ D,Z\right]  &= - 2 Z \,, \nn\\
\left[ K, P_{a} \right]  &= - \e_{ab} G^{b} \,,          &  \left[ K, H\right] &= -  D \,,          &  \left[ K, Y \right] &= S \,, \nonumber\\
\left[ K, Z \right]  &=2 Y \,,        &  \left[ H, S \right]  &= - 2 Y \,, & \left[ H, Y \right]  &= - Z \,, 
\label{SchBoson}
\end{align}
The corresponding  bi-linear form invariant under the extended Schr\"odinger algebra is given by \cite{Hartong EBG}
\bea
&&(P_a,G_b) = \delta_{ab}\,, \quad (H,S) =-1 \,, \quad (J,M) =-1 \,, \quad  (D,Y) = -1 \,, \quad (K,Z) =  -1\,, \label{BosonicSchrodingerMetric}
\eea
Taking the gauge field to be
 \bea
A_\m &=& \t_\m H + e_\m{}^a P_a + \o_\m J + \o_\m{}^a G_a + m_\m M + s_\m S +  b_\mu D + f_\m K + y_\mu Y +z_\m Z  \,,
\label{GFBosSch}
\eea
we can construct the extended Schr\"odinger gravity action \cite{Hartong EBG}
\bea
S &=&\frac{k}{4 \pi} \int d^3x\, \ve^{\m\n\r}  \Big( \o_\m{}^a R_{\n\r a} (P) - 2 s_\m R_{\n\r} (H) - 2 m_\m R_{\n\r}(J) + e_\m{}^a R_{\n\r a} (G) \nn\\
&&  \quad  - 2 y_\m R_{\n\r} (D) - 2 z_\m R_{\n\r} (K)    \Big) \,.
\label{BosonicSchrodinger}
\eea
The supersymmetric extension of the extended Schr\"odinger algebra requires the inclusion of two new supercharges, $F^\pm$ that are both Majorana spinors. With the assistance of the computer algebra program \textit{Cadabra} \cite{Cadabra1,Cadabra2}, we found that the non-vanishing $\left[B,B\right]$ commutators of the extended Schr\"odinger superalgebra are given by (\ref{SchBoson}) while the non-vanishing $\left[B,F\right]$ commutators are
\begin{align}
[J, Q^\pm ] & = - \frac{1}{2} \g_0 Q^\pm \,, & [J, R ] &= - \frac{1}{2} \g_0 R  \,, & [J, F^\pm ] & = - \frac{1}{2} \g_0 F^\pm  \,,\nn\\
[S, Q^+ ] &= - \frac{1}{2} \g_0 R \,, & [G_a, Q^+  ] & = - \frac{1}{2} \g_a Q^- \,, & [G_a, Q^-  ] &= - \frac{1}{2} \g_a R \,,\nn\\
[P_a, F^+] &= -\frac{1}{2} \epsilon_{ab} \g^b Q^-\,, & [P_a,Q^-] &= -\frac{1}{2} \epsilon_{ab} \g^b F^- \,, & [U_1, Q^\pm ] &= \pm  \g_0 Q^\pm \,,\nn\\
[U_1, R ] &=    \g_0 R \,,  &[U_1, F^\pm] &=    \g_0 F^\pm \,, &[U_2, Q^+ ] &=  - \frac{3}{4}   F^- \,,\nn\\
[U_2, F^+ ] &=   \frac{3}{4}  R \,, & [H, F^+] &=  - Q^+ \,, & [H, R ] &=    F^- \,, \nn\\
[D, Q^+] &= - Q^+ \,, & [D, F^\pm] &=  \pm F^\pm \,, & [D, R] &=  R \,,\nn\\
[K,Q^+] &= F^+ \,,  & [K,F^-] &= - R \,, &[M,Q^+] &= \frac{1}{4} F^- \,,\nn\\
 [M,F^+] &=  - \frac{1}{4} R \,, &[Y,Q^+] &= \frac{1}{4} \g_0  F^- \,, & [Y,F^+] &=  \frac{1}{4} \g_0  R \,,\nn\\
 [Z,F^+] &= -  \frac{1}{2} \g_0   F^-  \,,
\end{align}
while the non-vanishing $\{F,F\}$ anti-commutators are given by
\bea
&&\{ Q^+_\a ,  Q^+_\b  \} = (\g_0 C^{-1})_{\a \b} H \,,\nonumber\\
&&\{ Q^+_\a ,  Q^-_\b  \}  = -(\g_a C^{-1})_{\a \b} P^{a} ,,\nonumber\\
&&\{ Q^-_\a ,  Q^-_\b  \}  = (\g_0 C^{-1})_{\a \b} M  - (\g_0 C^{-1})_{\a \b} U_2\,, \nn \\
&&\{ Q^+_\a ,  R_\b  \}  = (\g_0 C^{-1})_{\a \b} M + 2 C_{\alpha \beta} Y + (\g_0 C^{-1})_{\a \b} U_2\,, \nn \\
&&\{ Q^+_\a ,  F^-_{\b}  \}  = - 2 C_{\alpha \beta} Z \,, \nn \\
&&\{ F^+_\a ,  F^+_\b  \} = (\g_0 C^{-1})_{\a \b} K \,,\nonumber\\
&&\{ F^+_\a ,  Q^+_\b  \} =  - \frac{1}{2}(\g_0 C^{-1})_{\a \b} D + \frac{1}{2} C_{\a\b} J  + \frac{3}{4} C_{\a \b} U_1 \,,\nonumber\\
&&\{ F^+_\a ,  Q^-_\b  \} =  \epsilon_{ab}(\g^a C^{-1})_{\a \b} G^b\,, \nn \\
&&\{ F^+_\a ,  R_\b  \} =  2 C_{\a \b} S\,, \nn \\
&&\{ F^+_\a ,  F^-_{\b}  \} =  (\g_0 C^{-1})_{\a \b} M - 2  C_{\a \b} Y + (\g_0 C^{-1})_{\a \b} U_2 \,.
\label{3dSchrodingerExtendedBargmann}
\eea
The supersymmetric extension of the extended  Schr\"odinger algebra admits the following invariant bilinear forms 
\bea
&&(P_a,G_b) = \delta_{ab}\,, \quad (H,S) =- 1\,, \quad (J,M) =-1 \,, \quad  (D,Y) = -1 \,,  \quad (K,Z) =  -1\,, \nn \\
&&(R_1,R_2) = 2  \,, \quad (Q^+_\alpha,R_\beta)= 2 C_{\a \b}\,,\quad (Q^-_\alpha,Q^-_\beta)= 2 C_{\a \b}  \,,\quad (F^+_\a,F_\b^-)= 2 C_{\a \b}\,. \label{FermionicSchrodingerMetric}
\eea
Taking the gauge field to be
\bea
A_\m &=& \t_\m H + e_\m{}^a P_a + \o_\m J + \o_\m{}^a G_a + m_\m M + s_\m S +  b_\mu D + f_\m K + y_\mu Y +z_\m Z \nn\\
&&+r_{1\m} U_1 +r_{2\m} U_2 + \bar\p^+_\m Q^+   + \bar\p^-_\m Q^-   + \bar\rho_\m R +   \bar\phi^+_\m F^+   + \bar\phi^-_\m F^-  \,,
\label{GFSuperSch}
\eea
we obtain the following Chern-Simons action
\bea
S &=&\frac{k}{4 \pi} \int d^3x\, \ve^{\m\n\r}  \Big( \o_\m{}^a  R_{\n\r a} (P) + e_\m{}^a  R_{\n\r a} (G)  - 2 s_\m  R_{\n\r} (H)    \nn\\
&& \quad - 2 m_\m  R_{\n\r}(J) - 2 y_\m  R_{\n\r} (D)  - 2 z_\m R_{\n\r}(K) +   4 r_{2\m} R_{\n\r}(U_1)  \nn\\
&& \quad +  4  \bar\r_\m R_{\n\r} (Q^+) + 2 \bar\p_\m^- R_{\n\r} (Q^-)  + 4 \bar\phi_\m^- R_{\n\r} (F^+) \Big) \,.
\label{SuperSch}
\eea
We refer to this action as the extended Schr\"odinger supergravity. The Schr\"odinger supercovariant curvatures are given as
\bea
R_{\m\n}(H) &=& 2 \partial_{[\m} \t_{\n]}  - 4 b_{[\m} \t_{\n]} + \bar\p_\m^+ \g_0 \p_\n^-   \,,\nn\\ 
R_{\m\n}{}^a (P) &=&  2 \partial_{[\m} e_{\n]}{}^a + 2 \e^{ab}\,  \o_{[\m} e_{\n]b} - 2 \e^{ab}\,  \o_{ [\m b} \t_{\n]}  - 2 b_{[\m} e_{\n]}^a- 2 \bar\p_{[\m}^+ \g^a \p_{\n]}^-  \,,\nn\\ 
R_{\m\n}(M) &=& 2 \partial_{[\m} m_{\n]} + 2 \e_{ab} \o_{[\m}{}^a e_{\n]}{}^b + 2 \bar\p_{[\m}^+ \g_0 \r_{\n]} + \bar\p_\m^- \g_0 \p_\n^-  + 2 \bar\phi_{[\m}^+ \g_0 \varphi_{\n]} \,,\nn\\ 
R_{\m\n}{}^a (G) &=& 2 \partial_{[\m} \o_{\n]}{}^a + 2  \epsilon{}^{ab} \o_{[\m}{} \o_{\n] b}  + 2  \epsilon{}^{ab} f_{[\m}{} e_{\n] b} + 2 b_{[\m} \o_{\n]}^a  - 2\epsilon^{a b} \bar \phi_{[\m}^+ \g_b \p_{\n]}^- \,,\nn\\
R_{\m\n} (J) &=& 2 \partial_{[\m} \o_{\n]} +  \bar\phi_{[\m}^+  \p_{\n]}^- \,,\nn\\
R_{\m\n} (S) &=& 2 \partial_{[\m} s_{\n]}  +  \epsilon{}^{ab} \o_{[\m a} \o_{\n]b }   +  4 b_{[\m} s _{\n]} + 2 f_{[\m} y_{\n]}+ 4 \bar\phi^+_{[\m}  \r_{\n]} \,, \nn \\
R_{\m\n} (K) &=& 2 \partial_{[\m} f_{\n]}   +  4 b_{[\m} f _{\n]} +  \bar\phi^+_{\m} \g_0  \phi^+_{\n} \,, \nn \\
R_{\m\n} (D) &=& 2 \partial_{[\m} b_{\n]}   +  2 \t_{[\m} f _{\n]}  -  \bar\phi_{[\m}^+ \g_0 \p_{\n]}^+ \,, \nn \\
R_{\m\n} (Y) &=& 2 \partial_{[\m} y_{\n]}  -  2  \o _{[\m a} e_{\n] }^a +  4 f_{[\m}  z_{\n]} - 4 \t_{[\m} s _{\n]} +  4 \bar\p_{[\m}^+  \r_{\n]} - 4 \bar\phi_{[\m}^+  \phi^-_{\n]} \,, \nn \\
R_{\m\n} (Z) &=& 2 \partial_{[\m} z_{\n]}  +  \epsilon{}^{ab} e_{[\m a} e_{\n]b }  - 4  b_{[\m} z_{\n]} - 2 \t_{[\m} y _{\n]} - 4 \bar\p_{[\m}^+  \varphi_{\n]}\,, \nn \\
R_{\m\n}(U_1) &=& 2 \partial_{[\m} r_{1\n]}   +  \frac{3}{2}\bar\phi^+_{[\m}  \p_{\n]}^+ \,,\nn\\ 
R_{\m\n}(U_2) &=& 2 \partial_{[\m} r_{2\n]}  +  2 \bar\p_{[\m}^+ \g_0 \r_{\n]} -  \bar\p_\m^- \g_0 \p_\n^-  +  2 \bar\phi_{[\m}^+ \g_0 \varphi_{\n]} \,,\nn \\
R_{\m\n}(Q^+) &=& 2 \partial_{[\m} \p_{\n]}^+ +  \o_{[\m} \g_0 \p_{\n]}^+ - 2 \t_{[\m} \phi^+_{\n]}  - 2 b_{[\m} \p_{\n]}^+  -  2 r_{1[\m} \g_0 \p_{\n]}^+  \,,\nn\\ 
R_{\m\n}(Q^-) &=& 2 \partial_{[\m} \p_{\n]}^- +  \o_{[\m} \g_0 \p_{\n]}^-  + \o_{[\m}{}^a \g_a \p_{\n]}^+  + \epsilon^{a b} e_{[\m a} \g_b \phi_{\n]}^+  +    2   r_{1[\m} \g_0 \p_{\n]}^-  \,,\nn\\ 
R_{\m\n}(R) &=& 2 \partial_{[\m} \rho_{\n]} +  \o_{[\m} \g_0 \rho_{\n]}  + \o_{[\m}{}^a \g_a \p_{\n]}^- +  s_{[\m} \g_0 \p_{\n]}^+  -   2 r_{1[\m} \g_0 \rho_{\n]} - \frac{3}{2} r_{2[\m}  \phi^+_{\n]}  + 2 b_{[\m}  \r_{\n]} \nn \\
&&  - 2 f_{[\m}  \phi^-_{\n]}  - \frac{1}{2} m_{[\m} \phi^+_{\n]} -   \frac{1}{2} y_{[\m} \phi^+_{\n]}  \nn \\ 
R_{\m\n}(F^+) &=& 2 \partial_{[\m} \phi^+_{\n]} +  \o_{[\m} \g_0 \phi^+_{\n]}  -   2 r_{1[\m} \g_0 \phi^+_{\n]} + 2 b_{[\m}  \phi^+_{\n]}  +  2 f_{[\m}  \p_{\n]}^+   \nn \\ 
R_{\m\n}(F^-) &=& 2 \partial_{[\m} \phi^-_{\n]} +  \o_{[\m} \g_0 \phi^-_{\n]}   -   2 r_{1[\m} \g_0 \phi^-_{\n]}  - 2 b_{[\m}  \phi^-_{\n]} +  \frac{3}{2} r_{2[\m}  \p_{\n]}^+  +  2 \t_{[\m}  \r_{\n]}   \nn \\
&&  + \frac{1}{2} m_{[\m} \p_{\n]}^+ -   \frac{1}{2} y_{[\m} \p_{\n]}^+  + z_{[\m}  \g_ 0 \phi^+_{\n]} \,. \nn  \label{Conformalsupercurvatures}
\eea
Finally, the extended Schr\"odinger supergravity action is invariant under the following transformation rules 
\bea
\d \t_\m &=&  - \bar{\e}^+ \g_0 \p^+_\mu \,,\nn\\
\d e_\m{}^a &=&    \bar{\e}^+ \g^{a} \p_\m^- +  \bar{\e}^- \g^{a} \p_\m^+ \,,\nn\\
\d \o_\m{}^a &=&  \epsilon^{ab} \bar{\e}^- \g_{b} \phi^+_\m +  \epsilon^{ab} \bar{\zeta}^+ \g_{b} \p_\m^- \,,\nn\\
\d \o_\m &=& \frac{1}{2} \bar{\e}^+  \phi^+_\mu -\frac{1}{2} \bar{\zeta}^+ \p_\mu \,,\nn\\
\d m_\m &=& - \bar{\e}^- \g_0 \p_\m^- -  \bar{\e}^+ \g_0 \rho_\m -  \bar{\eta} \g_0 \p_\m^+   -  \bar{\zeta}^+ \g_0 \phi^-_\m -  \bar{\zeta}^- \g_0 \phi^+_\m\,,\nn\\
\d s_\m  &=&  2 \bar{\eta} \phi^+_\m - 2 \bar{\zeta}^+ \r_\m \,,\nn \\
\d r_{1\m}  &=&   \frac{3}{4}\bar{\e}^+  \phi^+_\mu -  \frac{3}{4}\bar{\zeta}^+  \p_\mu^+ \,, \nn \\
\d r_{2\m}  &=&   \bar{\e}^- \g_0 \p_\m^- -  \bar{\e}^+ \g_0 \rho_\m -  \bar{\eta} \g_0 \p_\m^+   -  \bar{\zeta}^+ \g_0 \phi^-_\m -  \bar{\zeta}^- \g_0 \phi^+_\m\,,\nn\\
\d b_\m  &=&  \frac{1}{2} \bar{\e}^+ \g_0 \phi^+_\m  +  \frac{1}{2} \bar{\zeta}^+ \g_0 \p_\m^+\,,\nn \\
\d f_\m &=&  - \bar{\zeta}^+ \g_0 \phi^+_\mu \,,\nn\\
\d y_\m  &=&  2 \bar{\eta} \p_\m^+ - 2 \bar{\e}^+ \r_\m -2 \bar{\zeta}^- \phi^+_\m + 2\bar{\zeta}^+ \phi^-_\m \,,\nn \\
\d z_\m  &=&  -2 \bar{\zeta}^- \p_\m^+ +  2 \bar{\e}^+ \phi^-_\m \,,\nn\\
\d \p_\m^+ &=& \partial_\m \e^+ + \frac12 \o_\m \g_0 \e^+ -  \t_\m {\zeta}^+  - b_\m \e^+ -  r_{1\m} \g_0 \e^+ \,,\nn\\
\d \p_\m^- &=&  \partial_\m \e^- + \frac12 \o_\m \g_0 \e^- + \frac12 \o_\m{}^a \g_a \e^+  +  \frac{1}{2} \epsilon^{ab} e_{\m a} \g_b {\zeta}^+  +   r_{1\m} \g_0 \e^- \,,\nn\\
\d \r_\m &=&  \partial_\m \eta + \frac12 \o_\m \g_0 \eta + \frac12 \o_\m{}^a \g_a \e^- + \frac12 s_\m \g_0 \e^+  - \frac{1}{4}  m_\m {\zeta}^+  \nn\\
&& -  r_{1\m} \g_0 \eta  +  \frac{3}{4}  r_{2\m} {\zeta}^+   + b_\m \eta - f_\m {\zeta}^- - \frac{1}{4} y_\m \g_0 {\zeta}^+   \,, \nn\\
\d \phi_\m^+ &=& \partial_\m \zeta^+ + \frac12 \o_\m \g_0 \zeta^+   +  b_\m \zeta^+  + f_\m \e^+-  r_{1\m} \g_0 \zeta^+ \,,\nn\\
\d \phi^-_\m &=&  \partial_\m {\zeta}^- + \frac12 \o_\m \g_0 {\zeta}^- + \frac12 \e_{ab} e_\m{}^a \g^b \e^-   + \frac{1}{4}  m_\m \e^+ +   r_{1\m} \g_0 {\zeta}^- \nn\\
&& - \frac{3}{4}r_{2\m} \e^+  +  \frac{1}{2}  z_{\m} \g_0 {\zeta}^+  -  b_\m {\zeta}^- +  \t_\m \eta -  \frac{1}{4} y_\m \g_0 \e^+  \,,
\label{GaugeSchrodingerSusy} 
\eea
where $\zeta^\pm$ are the the parameters of the local $F^\pm$ transformations, respectively. Due to the map between the Newton-Cartan geometry and Horava-Lifshitz gravity \cite{Horava}, our action (\ref{SuperSch}) correspond to a superconformal non-projectable Horava-Lifshitz gravity \cite{Hartong EBG}.

\section{Discussion}

In this paper, we establish the supersymmetric extension of the extended Newton-Hooke, Lifshitz and Schr\"odinger algebras and construct the corresponding Chern-Simons supergravity models. The extended Newton-Hooke superalgebra admits two distinct non-degenerate invariant bi-linear form that gives rise to two different supergravity models with the same equations of motion. These two models are particularly different in terms of the parity of the bosonic actions. In particular, we showed that there is an exotic non-relativistic model such that parity-even field equations arise from a parity-odd Lagrangian. We then showed that it is possible to improve the extended Bargmann superalgebra with dilatations (without including non-relativistic special conformal symmetry) which we called the extended Lifshitz superalgebra and also established the Chern-Simons extended Lifshitz supergravity action. In the final step, we include the non-relativistic special conformal symmetry and establish the extended Schr\"odinger superalgebra and the corresponding Chern-Simons extended Schr\"odinger supergravity action.

We consider our paper as a first step to construct an off-shell formulation for the extended Bargmann supergravity and its matter couplings. Therefore, it would be very useful to study supermultiplet representations of the extended Schr\"odinger superalgebra. In particular, based on the bosonic construction \cite{AfsharBergshoeff}, it is natural to expect that a multiplet with a complex scalar field $\P$ as the lowest element can be used to construct a super-Schr\"odinger invariant gravity model. Taking multiple number of such multiplets and gauge fixing the Schr\"odinger supergravity action would give rise to the supergravity coupling of such a multiplet to off-shell supergravity, which is an important step towards non-relativistic localization. 

The $\cN=2$ supergravity models that we established here can be extended to higher number of supercharges. In this case, the extended Bargmann superalgebra with $\cN > 2$ can be obtained by Lie algebra expansion of the relativistic three-dimensional superalgebras. However, the Lifshitz and the Schr\"odinger superalgebras cannot be accessed by an Inonu-Wigner contraction or a Lie algebra expansion. In particular, given that the Schr\"odinger algebra is a particular instance $(\ell=1/2)$ of the so called $\ell$-conformal Galilei algebra \cite{Negro,Henkel}, in which case its $\cN > 2$ supersymmetric extensions exist \cite{Galajinsky1,Galajinsky2}, it would be very nice to initiate a systematic study to establish extended Schr\"odinger superalgebras for $\cN>2$ supersymmetry. There are also other three-dimensional non-relativistic algebras with an action principle such as the extended Newtonian supergravity \cite{Ozdemir}. Therefore, such a systematic study would also be helpful to uncover the Schr\"odinger extension of other three-dimensional superalgebras and hopefully towards a catalog of non-relativistic supergravity, at least in three dimensions. As a final remark, we would like to point out an interesting relation between the extended Schr\"odinger algebra and an extended Poincar\'e algebra. The extended Schr\"odinger algebra (and extended $\ell$-conformal Galilean algebras) has a hidden relativistic structure such that they can be written in a manifestly 3d  Lorentz-covariant form \cite{Sorokin}. It might of interest to see if the superextension of the Schr\"odinger superalgebra and the corresponding supergravity theory that we constructed in this paper can also be recast in a manifestly relativistic form.

\section*{Acknowledgements}

We thank Eric Bergshoeff for useful comments and suggestions. The authors would like to thank the referee for insightful comments which helped to improve this paper. The work of M.O. is supported in part by TUBITAK grant 118F091. N.O and U.Z. are supported in parts by Istanbul Technical University Research Fund under grant number TDK-2018-41133.
 
\appendix

\section{Comparison Between Different Notations} \label{AppB}

In this paper, we use the conventions of \cite{Bergshoeff EBG}. In what follows we introduce the  correspondence between the notations of \cite{EBG1}, \cite{Bergshoeff EBG} and \cite{Hartong EBG} for reader’s convenience.

\begin{table}[h!]
\centering
\begin{tabular}{|| c|| c || c ||}
 \hline \hline
Bergshoeff et. al \cite{Bergshoeff EBG} & Papageorgiou et. al. \cite{EBG1}& Hartong et. al. \cite{Hartong EBG}\\ \hline \hline
$H$               &          $-H$         &         $-H$           \\
$P_a$             &            $-P_a$        &            $P_a$      \\
$G_a$             &         $\e_{ab} K^b$            &              $\e_{ab} G^b$    \\
$J$               & $J$                  & $-J$            \\
$M$               &           $-M$           &        $-N$         \\
$S$               &         $S$             &               $S$ \\ \hline \hline
\end{tabular}
\end{table}

\section{Extended $\cN = (2,0)$ Newton-Hooke Superalgebra from Lie Algebra Expansion} \label{AppA}

In this appendix, we briefly discuss the Lie algebra expansion origin of the extended Newton-Hooke superalgebra (\ref{CosmoEBG1}) and (\ref{CosmoEBG2}). Lie algebra expansion, which was originally formulated in \cite{LAE1} and further studied in \cite{LAE2,LAE3}, is a methodology to generate larger Lie algebras starting from a specific one. To obtain larger algebras one first writes down a series expansion for Maurer-Cartan one-forms in powers of an expansion parameter $\l$, then inserts these expansions into Maurer-Cartan equation. Finally, the infinite expansions for the Maurer-Cartan form is cut in a consistent way from which one reads off the structure constants of the larger Lie algebra by matching the two sides of the Maurer-Cartan equation in powers of $\l$. We refer \cite{LAE4} to readers interested in such three-dimensional bosonic examples in more detail.

The algebra we shall consider here is the $\cN=(2,0)$ AdS superalgebra, which consists of the generators of space-time translations $P_A$, Lorentz transformations $J_A$ $(A = 0,1, 2)$, $R$-symmetry generator $U$ and two supersymmetry generators $Q^{1,2}$ that are both Majorana spinors. The non-zero $[B,B]$ and $[B,F]$ commutation relations for this algebra read \cite{3DAdS1,3DAdS2}
\begin{align}
\left[ J_{A}, J_{B} \right] &= \epsilon_{ABC} J^C\,, & \left[ J_{A}, P_{B} \right] &= \epsilon_{ABC} P^C \,, &\left[ P_{A}, P_{B} \right] &= \frac{1}{\ell^2}\ \epsilon_{A B C} J^C\,, \nn\\
[J_{A}, Q^{1,2}] &=  \frac{1}{2} \g_{A} Q^{1,2}\,, & [P_{A}, Q^{1,2}] &= \frac{1}{2 \ell}\ \g_{A}  Q^{1,2} \,, & [U, Q^{1,2}] &=  \pm  \frac{1}{\ell}\ Q^{2,1} \,,
\end{align}
while the non-vanising $\{F,F\}$ anti-commutators are given by
\bea
&& \{ Q^1_\a ,  Q^1_\b  \} =  2 (\g_A C^{-1})_{\a \b} P^A  + \frac{2}{\ell}\ (\g_A C^{-1})_{\a \b}J^A\,, \nn \\
&& \{ Q^1_\a ,  Q^2_\b  \} = 2 C_{\a \b} U\,,\nn \\
&& \{ Q^2_\a ,  Q^2_\b  \} =  2 (\g_A C^{-1})_{\a \b} P^A  +  \frac{2}{\ell}\ (\g_A C^{-1})_{\a \b} J^A\,. 
\eea
To implement the Lie algebra expansion method, we follow the methodology introduced in \cite{LAE4,LAE5,LAE6, LAE7} and decompose the 3-dimensional index $A$ as $A = (0,a)$ in which case the we have
\bea
J_A = (J, G_a)\,, \qquad P_A = (H, P_a) \,.
\eea
Furthermore, we choose the following combitation of the fermionic generators,
\bea
\tilde{Q}^\pm= \frac{1}{2}\big(Q^1 \pm  \g_0 Q^2\big)\,.
\eea
We, then, introduce the following gauge fields associated to the generators 
\begin{align}
J &\rightarrow \Omega\,, & G_a &\rightarrow \O^a \,, & H &\rightarrow \theta \,, & P_a &\rightarrow E^a \,,\nn\\
U & \rightarrow A \,, & \tilde{Q}^+& \rightarrow {\Psi}^+  \,,  & \tilde{Q}^-& \rightarrow {\Psi}^-  \,
\end{align}
The expansion of the gauge fields are given by
\begin{align}
\O &= \o + \l^2 s \,, & E^a &= \l e^a \,, & \theta &= \t+ \l^2 m \,, &\O^a &= \l \o^a  \,,\nn\\
A & = r_1 + \l^2 r_2 \,, & \P^+ &= \p^+ + \l^2 \r \,, & \P^- &= \l \p^- \,.
\end{align}
Using these expansions into the prescription given above, we precisely arrive at the extended Newton-Hooke superalgebra (\ref{CosmoEBG1}) and (\ref{CosmoEBG2}).

\end{document}